\documentclass[aps,prd,twocolumn,10pt,showpacs]{revtex4-2}
\usepackage[colorlinks=true, citecolor=blue, urlcolor=red]{hyperref}
\usepackage[utf8]{inputenc}
\usepackage{graphicx}
\usepackage{subfigure}
\usepackage{setspace}
\usepackage{amsmath}
\usepackage{commath}
\usepackage{amsthm}
\usepackage{color}
\usepackage{color,soul}
\usepackage{appendix}
\usepackage{tgpagella}
\usepackage{amssymb}
\allowdisplaybreaks
\usepackage[dvipsnames]{xcolor}
\begin{document}

\title{Probability distribution for black hole evaporation}

\author{Pratik Ghosal}
\email{ghoshal.pratik00@gmail.com}
\affiliation{Department of Physics,
Bose Institute, Unified Academic Campus, EN 80, Sector V, Bidhan Nagar, Kolkata 700091, India}

\author{Rajarshi Ray}
\email{rajarshi@jcbose.ac.in}
\affiliation{Department of Physics,
Bose Institute, Unified Academic Campus, EN 80, Sector V, Bidhan Nagar, Kolkata 700091, India}


\begin{abstract}
Non-thermal correction to the emission probability of particles from black
holes can be obtained if the backreaction or self-gravitational effects of the 
emitted particles on the black hole spacetime are taken into consideration.
These non-thermally emitted particles conserve the entropy of the black hole,
i.e, the entropy of the system of radiated particles after complete evaporation 
of the black hole matches the initial entropy of the black hole. Using the
non-thermal emission probability, we have determined the probability
for a black hole of mass $M$ to be completely evaporated by a given 
number of particles $n$.
This is done by first evaluating the number of possible ways in which the 
black hole can be evaporated by emitting $n$ number of particles, and then the total number 
of ways in which the black hole can be evaporated. The ratio of 
these two quantities gives us the desired probability. From the probability
distribution, we get a displacement relation between the most probable number 
of particles exhausting the black hole and the temperature of the initial 
black hole. This relation resembles Wien's displacement law for blackbody
radiation.
\end{abstract}

\maketitle

\section{Introduction}

Unifying general relativity and quantum mechanics - two important
pillars of modern physics, has been a long sought-after goal for
physicists. However, despite considerable efforts, a complete theory of 
quantum gravity remains elusive. A relatively simpler approach to studying 
quantum phenomena in gravitational fields is to treat matter fields
quantum mechanically and background spacetimes classically. This is a 
semi-classical approach, much like studying atomic physics in classical
electromagnetic fields instead of using full quantum electrodynamics. An
important landmark in this line of research is the discovery of Hawking
radiation by S.W. Hawking\cite{Hawking:1974sw}. Hawking showed that 
quantum effects of matter fields in the vicinity of black holes lead 
to the creation of particle-antiparticle pairs. The antiparticle tunnels 
through the event horizon, inside the black hole, and the particle is 
emitted outside. Thus, black holes act like hot bodies with temperature
$T_{BH}=\frac{1}{8\pi GM}$, emitting radiation. The discovery of Hawking
radiation not
only predicts a completely revolutionary phenomenon - i.e., emission of
particles from black holes which were classically thought to be
``regions of no return" - but it also poses some deep questions about the
nature of the interplay between quantum mechanics and gravity.
Consider a black hole that may have been formed due to the gravitational 
collapse of a star. The
information about the quantum states of its forming matter, which has
crossed the event horizon, is not accessible to an observer outside the
event horizon. But the information is stored safely beyond the horizon
(until it reaches the singularity at the centre). Now with the
discovery of Hawking radiation, as the black hole emits thermal
radiation it evaporates completely, leaving behind no
trace of the information about its forming matter. On the other hand,
thermal radiation does not carry any information as it escapes
the black hole. Thus, information seems to be lost in the black hole
evaporation process \cite{Hawking:1976ra}. This loss of
information is in contradiction to the principle of ``unitary evolution" in
quantum mechanics. So, Hawking radiation presents a conflict of
principles between general relativity and quantum mechanics. As a result
of non-unitary evolution, entropy is not conserved in the evaporation
process \cite{Zurek:1982zz}. In particular, the entropy of the
radiation system obtained after complete evaporation of the black hole
appears to be more than the Bekenstein-Hawking 
entropy \cite{Bekenstein:1973ur} of the
initial black hole ($S_{rad}=\frac{4}{3}S_{BH}$). Over the past few decades, a
substantial amount of work has been done on the \textit{``Information Loss
Paradox"} and its possible resolution
\cite{Hawking:2005kf,Preskill:1992tc,PhysRevLett.71.3743,Aharonov:1987tp,Krauss:1988zc,Bekenstein:1993bg,Horowitz:2003he,Lloyd:2004wn,Braunstein:2006sj,Mathur_2009,PhysRevD.48.3743,almheiri2013black,maldacena2013cool,RevModPhys.93.035002,almheiri2019entropy,penington2020entanglement}.

An important work by Parikh and Wilczek\cite{Parikh:1999mf} showed that the radiation
spectrum from a black hole is not strictly thermal if the 
self-gravitational effects of the emitted
particles on the black hole spacetime are taken into
consideration. It has been further shown that the
evaporation of black holes by the emission of non-thermal radiation is
consistent with the principle of unitary evolution of quantum
mechanics \cite{Parikh:2004rh}. If the evolution is unitary, one
expects to get back all the information that was stored inside the
black hole from the emitted radiation. Indeed, Zhang \textit{et}
al. \cite{Zhang:2009jn} showed that non-thermally emitted particles
share correlations between them in the form of mutual information. These
correlations carry information out of the black hole, and one can get
back the entire information by collecting all of these particles. As a
result, entropy is conserved in the evaporation process. To show the
conservation of entropy in the black hole evaporation process, Zhang \textit{et}
al. have calculated the entropy of the system of radiation obtained after complete evaporation of the black hole, which matches
the Bekenstein-Hawking entropy of the initial black
hole. However, this resolution of the information loss paradox is not unanimously
accepted. As pointed out in Ref.\cite{mathur2011information}, although 
entropy is conserved in the non-thermal radiation process, it does not
account for the increase of entanglement entropy between the 
inside and outside of the black hole during the entire period of
evaporation, which is the main essence of
the information loss paradox. Also see Ref.\cite{cai2012comment}.
Very recently, a new resolution to the paradox has been proposed
in Refs.\cite{almheiri2019entropy,penington2020entanglement} which 
is in line with the objections raised in Ref.\cite{mathur2011information}.

However, whether or not these works resolve the paradox, we can still
get some further useful information regarding black hole evaporation from
them. A black hole of some given mass $M$ may be completely evaporated by the
emission of an arbitrary number of particles. The number of particles that are obtained after the complete evaporation of the black hole is uncertain. Building upon the earlier works on non-thermal radiation, we 
have evaluated the probability that the black hole evaporates completely by emitting a given number of particles $n$. To do so, first we have evaluated the number of 
possible ways in which the black hole can be evaporated by emitting $n$  
particles. Then, we have evaluated the total number of ways in which the black 
hole can be evaporated. The ratio of 
these two quantities gives us the desired probability. From the probability
distribution, we get a relation between the most probable number of particles
exhausting the black hole and the temperature of the initial black hole.

The outline of this article is as follows. In section \ref{2}, we 
briefly discuss non-thermal Hawking radiation. In particular, we  
discuss, following the works of Parikh \cite{Parikh:2004rh} and Zhang \textit{et} al.
\cite{Zhang:2009jn}, how non-thermal radiation is consistent with the
unitarity of quantum mechanics, and how these emitted particles leak
the information stored inside the black hole and conserve entropy.
In section \ref{3}, we calculate the probability for a black hole to be 
completely evaporated by a given number of particles that are emitted 
non-thermally. From the probability distribution obtained, we get a 
displacement relation between the most probable number of particles 
exhausting the black hole and the initial black hole temperature. Then, we try to interpret the meaning of the entropy of the radiation system, obtained after the complete evaporation of the black hole. 
Finally, we summarise our results and conclude in section \ref{4}.
Throughout the paper, we use natural units ($c=\hbar=k_B=1$)
unless otherwise mentioned.

\section{\label{2} Non-thermal Hawking Radiation}
The evolution of a black hole to a thermal state, irrespective of 
its initial state, is non-unitary in nature and does not
preserve information. In the original work \cite{Hawking:1974sw} on 
Hawking radiation,
Hawking considered a static spacetime geometry for a black hole which is 
not perturbed
by the loss of energy of the emitted particles. Since the spacetime geometry is
not perturbed during the emission of a 
particle with energy $E$, the mass parameter $M$ of the black
hole spacetime remains unchanged during the emission period of that particle. This
violates the principle of energy conservation.  
Parikh, Kraus, and Wilczek \cite{Parikh:1999mf,Kraus:1994by} considered the case of dynamic geometry to enforce energy conservation. The dynamic
nature of the geometry is due to the varying mass parameter of the black hole spacetime. 
It takes into account the backreaction of the
emitted particles on the spacetime. The calculation of the tunneling 
probability through the event horizon, of a particle from the inside of a black hole
to the outside
or equivalently, of an antiparticle from the outside to the inside, in this dynamic geometry,
results in 
non-thermal correction terms to Hawking's original calculation\cite{Parikh:1999mf}. The
modified tunneling or emission probability of a particle with energy $E$ from a black
hole of mass $M$, up to a constant factor, is given
by
\begin{eqnarray}
\label{non_thermal}
\Gamma (E;M) &\sim& \exp \left[-8 \pi GE\left(M-\frac{E}{2}\right)\right] \nonumber\\
&=& \exp \left[4\pi G\left((M-E)^2-M^2\right)\right] \nonumber\\
&=& \exp\left[\Delta S_{BH}\right],
\end{eqnarray}
where $S_{BH} = 4 \pi GM^2$ is the Bekenstein-Hawking entropy of the
black hole of mass $M$ \cite{Bekenstein:1973ur}. The first term in the exponent
corresponds to thermal emission,
whereas the second term gives the non-thermal correction. The constant prefactor
of the exponential term in Eqn.(\ref{non_thermal}) can be determined by
including higher orders of $\hbar$ corrections \cite{Singleton:2010gz} (quantum
corrections) to the tunneling calculation done by Parikh and
Wilczek \cite{Parikh:1999mf}.

In Ref.\cite{Parikh:1999mf}, the emission probability is estimated by considering
modes of emitted particles that propagate from a point that is arbitrarily close to the
horizon of the black hole. Due to infinite gravitational redshift near
the black hole horizon, the modes can have an arbitrarily small wavelength
($\lambda \rightarrow 0$). However, there are indications from many
theories of quantum gravity of the existence of an observer-independent
minimum length which is identified with the Planck length
($l_P=\sqrt{\frac{\hbar G}{c^3}}$). Using the Planck length cut to the
wavelength of the modes (quantum gravity correction), the spectrum is
recalculated by Arzano \textit{et} al.\cite{Arzano:2005rs}. This
calculation again gives the prefactor of Eqn.(\ref{non_thermal}).
However, this prefactor is quite obviously different from Ref.
\cite{Singleton:2010gz}, because two different mechanisms
are used. Both of these calculations also result in a
logarithmic correction term to the entropy of the black
hole 
\footnote{Logarithmic corrections to black hole entropy are also introduced
by string theory and loop quantum gravity calculations. While the coefficient $\alpha$ is
predicted to be negative by loop quantum
gravity \cite{Ghosh:2004rq}, its sign depends on the number of
field species in the low energy approximation \cite{Solodukhin:1997yy} in
string theory.}.

\begin{equation}
S=\frac{A}{4G}+\alpha \ln \frac{A}{G},
\end{equation}
where $A=16\pi G^2 M^2$ is the area of the event horizon of the black hole.
\subsection{Non-thermal radiation is consistent with the principle of unitary evolution in quantum
mechanics}

In Ref.\cite{Parikh:2004rh}, it is argued that the evaporation process of a
black hole by the emission of non-thermal radiation is unitary in nature. In
quantum mechanics, the rate of a unitary process from an initial state $i$
to final state $f$ is given by
\begin{equation}
\Gamma(i \rightarrow f ) = |\mathcal{M}_{fi}|^2 \times
\text{(phase space factor)},
\end{equation}
where $\mathcal{M}_{fi}$ is the amplitude of the process. The phase
space factor is obtained by summing over all possible final states,
which is simply the exponential of the final entropy ($S_f$) of the
system, and averaging over all possible initial states, which is the
exponential of the initial entropy ($S_i$) of the system. Therefore,
\begin{eqnarray}
\Gamma(i \rightarrow f ) &\sim & \frac{e^{S_f}}{e^{S_i}}, \nonumber \\
&=& e^{\Delta S}.
\end{eqnarray}
Since Eqn.(\ref{non_thermal}) matches the above expression, one can
say that the evaporation process is also unitary in nature. 

\subsection{Correlation between non-thermal Hawking quanta}

Considering the non-thermal emission probability (Eqn.(\ref{non_thermal})),
Zhang \textit{et} al.\cite{Zhang:2009jn} showed
that this non-thermal Hawking radiation can carry information out of
a black hole in the form of correlation between sequential emissions of quanta.
Consider the successive emission of two particles with energies $E_1$ and $E_2$
from a black hole of mass $M$. The emission
probability of the first particle with energy $E_1$ is given by
\begin{equation}
\Gamma(E_1;M)=\exp\left[-8 \pi GE_1\left(M-\frac{E_1}{2}\right)\right].
\end{equation}
Now, after the emission of the first particle, the black hole's mass has
reduced to $M-E_1$. The emission probability of the second particle of
energy $E_2$ from the black hole of reduced mass is given by
\begin{equation}
\Gamma(E_2;M-E_1)=\exp\left[-8 \pi GE_2\left(M-E_1-\frac{E_2}{2}\right)\right].
\end{equation}
One can see that the emission probability of the second particle depends
on the energy of the first particle. This implies the existence of statistical
correlations between the two emissions. The joint probability of the two
emissions is
\begin{align}
\Gamma(E_1,E_2)=\Gamma(E_1&;M)\Gamma(E_2;M-E_1) \nonumber \\
=\exp&\left[-8 \pi G\left(ME_1 -\frac{E_1^2}{2}+ME_2-E_1E_2\right.\right. \nonumber\\
&\left.\left.-\frac{E_2^2}{2}\right)\right]\nonumber \\
=\exp&\left[-8 \pi G(E_1+E_2)\left(M-\frac{E_1+E_2}{2}\right)\right] \nonumber \\
=\Gamma(&E_1+E_2;M),
\end{align} 
where $\Gamma(E_1+E_2;M)$ is the emission probability of
a single particle with energy $E_1+E_2$ from a black hole of mass $M$.

The correlation function for the two emissions is defined
as\cite{Zhang:2009jn}
\begin{equation}
\chi(E_1,E_2)= \ln \left( \frac{\Gamma(E_1+E_2;M)}
{\Gamma(E_1;M)\Gamma(E_2;M)}\right),
\end{equation}
where the numerator is the probability of the emission of two particles
with energy $E_1$ and $E_2$ \textit{simultaneously}, or of a single
particle with total energy $E_1+E_2$, from a black hole of mass $M$, 
and the denominator is the product
of the probabilities of emission of particles of energies $E_1$ and
$E_2$, each occurring \textit{independently} from a black hole of the same mass.
Since each particle is emitted independently, their emission probabilities 
do not depend on the energy of the other particle. Therefore,
\begin{align}
\Gamma(E_1;M) &= \exp\left[-8 \pi GE_2\left(M-\frac{E_1}{2}\right)\right], \nonumber\\
\Gamma(E_2;M) &= \exp\left[-8 \pi GE_2\left(M-\frac{E_2}{2}\right)\right], \nonumber\\
\text{and,~~}\chi(E_1,E_2) &= 8 \pi GE_1 E_2.
\end{align}
So there exists a non-zero correlation between the two
emissions. Zhang \textit{et} al. argued that this implies that radiation can carry information out
of the black hole in the form of correlations between sequential emissions.
Defining the entropy of the emitted particles
as\cite{Zhang:2009jn}

\begin{equation}
S(E)=- \ln \Gamma(E),
\label{single_entropy}
\end{equation}
we get,
\begin{eqnarray}
\chi(E_1,E_2) &=& \ln \Gamma(E_1+E_2;M) - \ln \Gamma(E_1;M) \nonumber \\
& & -\ln \Gamma(E_2;M)] \nonumber \\
&=& S(E_1)+S(E_2)-S(E_1,E_2) \nonumber \\
&=& I(E_1:E_2)
\end{eqnarray}
which is simply the mutual information shared between the two
particles. So the statistical correlation between the two emissions is
shared in the form of mutual information.

\subsection{Entropy conservation by non-thermal radiation}
Consider the following scenario:
A black hole of mass $M$ is evaporated completely by the emission of 2 particles
with energies $E_1$ and $E_2$.
Therefore, $E_1+E_2=M$. Now from Eqn.(\ref{single_entropy}), the entropy of the first particle is
\begin{eqnarray}
S(E_1)&=& - \ln \Gamma(E_1;M) \nonumber \\
&=& 8 \pi GE_1\left(M-\frac{E_1}{2}\right),
\end{eqnarray}
and the entropy of the second particle is
\begin{eqnarray}
S(E_2|E_1)&=& - \ln \Gamma(E_2;M-E_1) \nonumber \\
&=& 8 \pi GE_2\left(M-E_1-\frac{E_2}{2}\right),
\end{eqnarray}
where $S(E_2|E_1)$ is the conditional entropy of the second particle,
provided that the first particle is emitted with energy $E_1$. Since the
emission probability of the second particle is dependent (conditional)
on the first particle, the entropy of
the second particle is also conditional. Therefore, the total entropy of
the system is
\begin{eqnarray}
S_{rad}&=& S(E_1)+S(E_2|E_1) \nonumber \\
&=& 4 \pi GM^2 = S_{BH}.
\end{eqnarray}
So, the total entropy of the emitted particles, after complete evaporation 
of the black hole, is the same as the initial Bekenstein-Hawking entropy of the
black hole.

This holds, in general for any number of particles exhausting
the black hole i.e., if the black hole is evaporated completely by the emission of
$n$ particles, then
\begin{eqnarray}
S_{rad}&=& S(E_1)+S(E_2|E_1)+\ldots+S(E_n|E_1,E_2,\ldots,E_{(n-1)}) \nonumber \\
&=& 4 \pi GM^2 = S_{BH},
\end{eqnarray}
where, $E_1+E_2+E_3+\ldots+E_n=M$. Note that entropy conservation holds true
independent of the individual energies of the emitted particles, and hence
the sequence of emissions also.

Zhang \textit{et} al.'s work on
correlation and entropy conservation by non-thermal Hawking radiation has also been supported by Ref.\cite{Israel:2010gk}.
The correlation and conservation of entropy has also been studied for the
case where the quantum gravity correction is
taken\cite{Chen:2009ut,Zhang:2009td}. It has been shown
that for quantum-gravity-corrected emission probability, the logarithmic
corrected entropy of the black hole is conserved only when there is a 
black hole remnant.
That means that the black hole cannot be evaporated completely if
the quantum gravity correction is taken into account.
\section{\label{3} Probability distribution}
\subsection{Probability of complete black hole evaporation by emission of $n$ particles}
We have seen that, for non-thermal emissions, entropy is conserved 
irrespective of the number of particles emitted and the energies of 
the individual particles. However, how many 
particles will be emitted before the black hole is completely evaporated 
is not known for certain. That is, the black hole can be completely 
evaporated by emitting a single particle, or infinitely many particles.
We wish to 
determine the probability ($q_n$) that a black 
hole of some
given mass $M$ is completely evaporated by the emission of a given number of particles $n$. For this,
first we have to know in how many ways ($\Omega_n$) the
black hole evaporates completely by the emission of $n$ particles. We will
clarify what these different ways correspond to in a moment. Also, we have to know
the total number of ways ($\Omega_{total}$) in which the black hole evaporates, where
\begin{equation}
    \Omega_{total}=\sum_{n=1}^\infty \Omega_n.
\end{equation}
Then, we can define the probability as
\begin{equation}
q_n = \frac{\Omega_n}{\Omega_{total}}.
\end{equation}

To determine $\Omega_n$, we consider the system of $n$ emitted
particles of total energy $M$ (mass of the initial black 
hole), obtained after complete evaporation of the black hole.
If this system can be obtained in $\Omega_n$ different possible ways, 
we may say that $\Omega_n$ is the number of microstates of the system 
corresponding to the macrostate defined by total energy $M$. Furthermore,
since entropy is conserved irrespective of the individual energies and the number of emitted
particles, the total entropy of this system is $S_{BH}$ for all the $\Omega_n$ possible
microstates.
Therefore, the
macrostate of this system can be defined by macroscopic properties
like total energy, total entropy, and the number of particles
($M,S_{BH},n$). Note that we do not have a well-defined volume of the system.

Having defined the macrostate of the system, now we need to define
precisely what we mean by the microstates of the system - or, in 
other words, what the different possible ways of black hole 
evaporation by $n$ particles correspond to. Since the total 
energy of the system is a macroscopic property which is same for all
the microstates, one may consider that different possible partitions
of the total energy among the individual particles form the 
microstates of the system. This is typical for microcanonical 
systems. But the system that we have considered here is different, in the
sense that (i) the volume of the system does not define the macrostate,
and (ii) the entropy of the system defines the macrostate of the system.
Furthermore, one can show that the conservation of entropy naturally leads to
the conservation of energy in the evaporation process. That is,
\begin{align}
\sum_{i=1}^n S_i=S_{BH}\Rightarrow\sum_{i=1}^n E_i=M, 
\end{align}
where, $S_i$ and $E_i$ are the entropy and energy of the $i^{th}$
emitted particle, respectively. Here, by conservation of energy we
mean that the sum of the individual energies of the emitted particles
equals the mass of the initial black hole. However, the reverse cannot be shown. So, conservation of entropy is more 
fundamental for our system, and we are essentially
left with only two independent macroscopic properties ($S_{BH},n$)
that define the macrostate of the system. For this reason, we define
the microstates of the system to be the different possible partitions
of the total entropy $S_{BH}$ among the individual particles. That is, every possible set of individual entropies,
\begin{align}
    \left\lbrace S_i,i=1,2,\ldots,n \middle| \sum_{i=1}^n S_i=S_{BH}\right\rbrace,
\end{align}
forms the microstates of the system.

\begin{figure}[hbt]
\centering
\includegraphics[height=7cm,width=7cm]{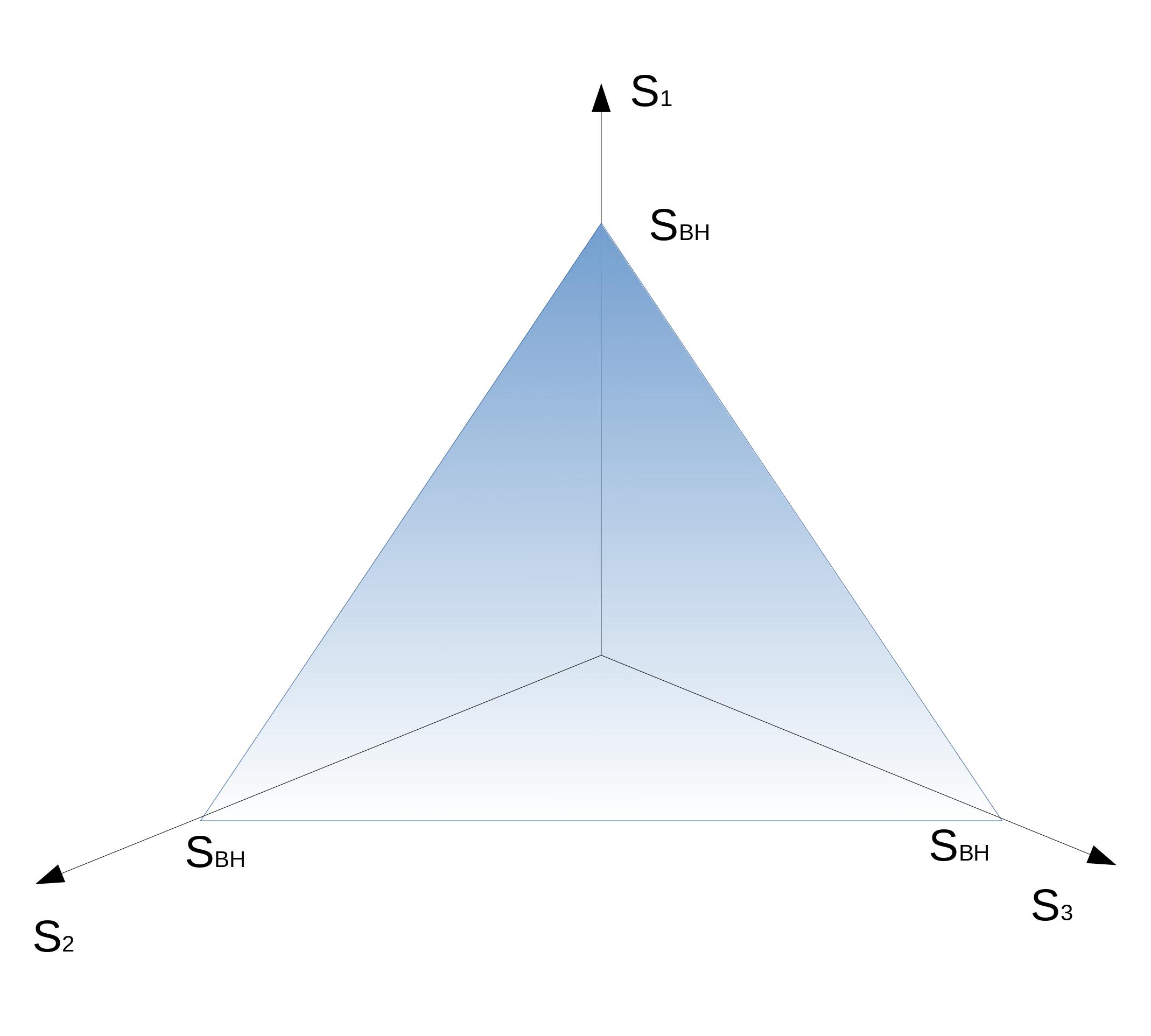}
\caption{\label{entropy_space} Three-dimensional entropy space for three-particle
evaporation of the black hole. The individual entropies of the emitted particles form the axes in the space. The shaded region is the microstate
hyperplane.} 
\end{figure}

Now, we need to evaluate the number of possible partitions of the total
entropy $S_{BH}$ to get $\Omega_n$. This can be done as follows.
Consider the \textit{entropy space} 
formed by entropy of the individual particles. Then, the equation
\begin{equation}
\label{constraint}
\sum_{i=1}^{n} S_i = S_{BH},
\end{equation} 
defines an ($n-1$)-dimensional hyperplane in the $n$-dimensional
entropy space. Moreover, since the entropy of any emitted particle is
always positive (see Eqn.(\ref{single_entropy})), the microstates
that we define, lie on the region of this hyperplane which is
bounded by the hyperplanes $S_i=0,\forall i=1,2,\ldots,n$.
The entropy space and the microstate hyperplane for three-particle
evaporation of the black hole are shown in Fig.(\ref{entropy_space}).
Now, the number of microstates can be given by
\begin{eqnarray}
& &\text{Number of microstates} (\Omega_n) \nonumber \\
&=& \frac{\text{Surface area of the microstate hyperplane}}
{\text{Area of a unit cell on the microstate hyperplane}} \nonumber \\
&=& \frac{\mathcal{S}_{n-1}}{\mathcal{A}_{n-1}},
\end{eqnarray}
where the unit cell effectively contains a single microstate.

The surface area of the ($n-1$)-dimensional microstate hyperplane defined
by Eqn.(\ref{constraint}) is given
by (See Appendix \ref{A})
\begin{equation}
\mathcal{S}_{n-1} = \sqrt{n} \frac{S_{BH}^{n-1}}{(n-1)!},
\end{equation}
and
the area of the unit cell on the microstate hyperplane is 
(See Appendix \ref{A})
\begin{equation}
\mathcal{A}_{n-1} = \sqrt{n}.
\end{equation}
\begin{figure}[!htb]
\centering
\includegraphics[height=7cm,width=8.5cm]{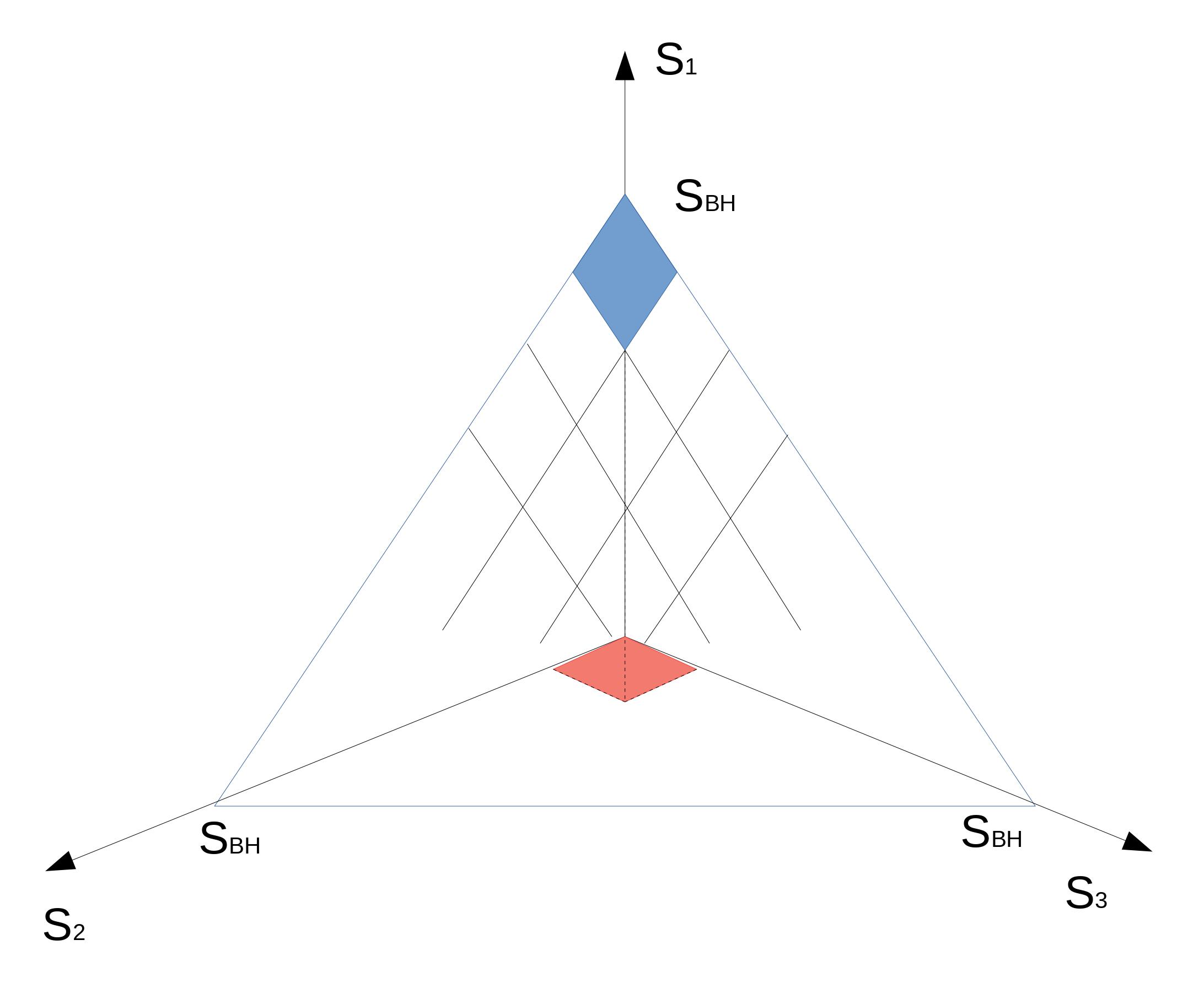}
\caption{\label{projection} Unit cell on the microstate hyperplane is shown
by the blue shaded region. It projects a unit square (two-dimensional cube)
on the space spanned by $S_2$ and $S_3$ (red shaded region). The area of the
unit cell is $\mathcal{A}_2=\sqrt{3}$.} 
\end{figure}
Therefore,
\begin{eqnarray}
\Omega_n &=& \frac{\sqrt{n} \frac{S_{BH}^{n-1}}{(n-1)!}}
{\sqrt{n}} \nonumber \\
&=& \frac{S_{BH}^{n-1}}{(n-1)!} \nonumber \\
&=& \frac{(4 \pi GM^2)^{n-1}}{(n-1)!}.
\end{eqnarray}
This gives us the number of different possible ways in which the black hole can 
be evaporated by the emission of $n$ particles. Now, the total number of ways in which the black hole
can evaporate is
\begin{eqnarray}
\label{omegatotal}
\Omega_{total} &=& \sum_{n=1}^\infty \Omega_n \nonumber \\
&=& \sum_{n=1}^\infty \frac{(4 \pi GM^2)^{n-1}}{(n-1)!} \nonumber \\
&=& e^{4 \pi GM^2} .
\end{eqnarray}

Therefore, the probability that the black hole is completely 
evaporated by the emission of $n$ particles is given by
\begin{eqnarray}
q_n &=& \frac{\Omega_n}{\Omega_{total}} \nonumber \\
&=& \label{q_n}\frac{(4 \pi GM^2)^{n-1}}{(n-1)!}e^{-4 \pi GM^2}.
\end{eqnarray}
It is easy to verify that the probabilities add up to unity,
\begin{eqnarray}
\sum_{n=1}^\infty q_n &=& e^{-4 \pi GM^2} \sum_{n=1}^\infty \frac{(4 \pi 
GM^2)^{n-1}}{(n-1)!} \nonumber \\
&=& e^{-4 \pi GM^2} \times e^{4 \pi GM^2} \nonumber \\
&=& 1.
\end{eqnarray}
That is, the probabilities, $q_n$ are normalised.
\subsection{Displacement relation}
If we plot the probabilities $q_n$ as a function of the numbers of
emitted particles $n$ from Eqn.(\ref{q_n}), we get the probability
distribution for the black hole to be evaporated completely by the emission
of different possible numbers of particles. The probability distribution 
is parametrised by the entropy of the black hole ($S_{BH}$).

\begin{figure}[htb]
\centering
\includegraphics[height=5cm,width=8cm]{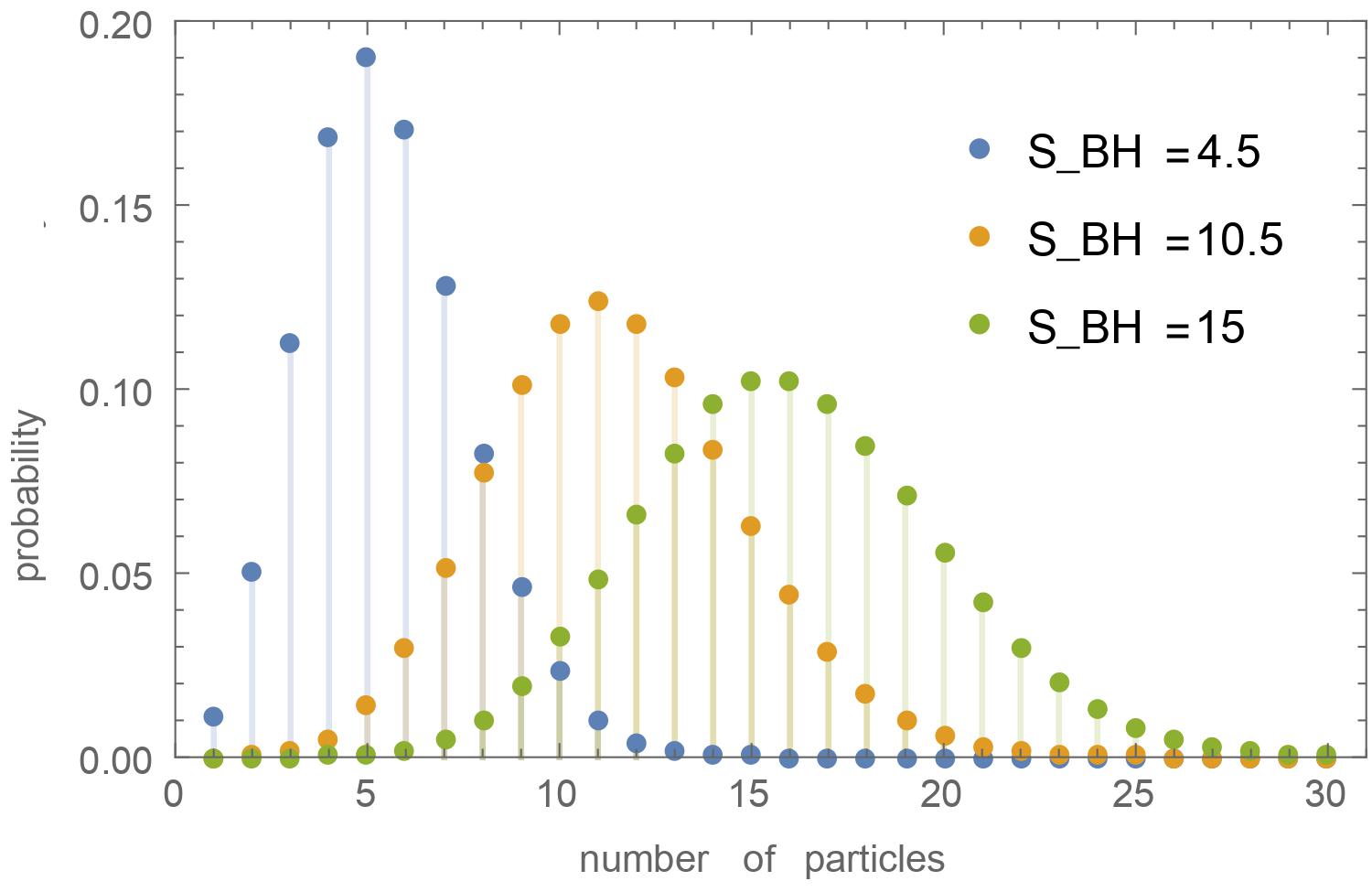}
\caption{\label{pd}Probability distribution for black holes of
different masses to be evaporated by the emission of different possible numbers of particles.} 
\end{figure}

From Fig.(\ref{pd}) we see that for a black hole of given mass, there exists
a certain number of particles, $n_{max}$, for which the probability
distribution peaks. That is, the black hole is most likely to
evaporate completely by the emission of $n_{max}$ particles.
We further see that, as the mass, and hence the entropy of the black hole
increases, the peak of the distribution decreases and shifts towards
higher values of $n_{max}$. For a black hole of given mass, $n_{max}$
has the nearest integer value between $S_{BH}$ and $S_{BH}+1$ (See Appendix \ref{B}), 
\begin{align}
S_{BH} &\leq n_{max} \leq S_{BH}+1,\\
\text{i.e,~~}n_{max} &\approx S_{BH} \nonumber\\
&= 4 \pi GM^2 \nonumber \\
&= \frac{1}{16 \pi GT_{BH}^2},
\end{align}
where, $T_{BH}$ is the Hawking temperature of the initial black hole.
In other words,
\begin{equation}
\label{displacement_relation}
n_{max} T_{BH}^2 = \frac{1}{16 \pi G} = \text{constant}.
\end{equation}

Eqn.(\ref{displacement_relation}) relates the most probable number of
emitted particles with the temperature of the initial black hole.
This relation resembles Wien's displacement law for blackbody
radiation. According to Wien's displacement law, the wavelength of blackbody radiation for which the spectral energy density is maximum, $\lambda_{max}$, is
inversely proportional to the temperature of the blackbody,
\begin{align}
\label{Wien}
\lambda_{max} T&=2.898\times 10^{-3}~\text{mK}~~~\text{(not in natural units)} \nonumber\\
&=\text{constant}.
\end{align}
\subsection{Entropy of the system}
We have seen how the conservation of entropy by the non-thermal emission of
particles from black holes plays an important role in determining
the possible number of ways in which a black hole can evaporate
completely by emitting a given number of particles. Before concluding,
let us delve a little more into the entropy of the radiation system,
or the system of emitted particles after complete evaporation of the
black hole. In information theory, the entropy of a system is considered as a ``measure
of uncertainty" of the system. If a system can exist in multiple
possible states, then there is an uncertainty about the state in which the system is.
For each possible state there is a probability which measures the likeliness of the
system to exist in that state. Given a probability distribution for the
system, we can quantify the uncertainty in terms of Shannon entropy as
\begin{equation}
S = - \sum_{i} p_i \ln p_i,
\end{equation}
where $p_i$ is the probability of the $i^{th}$ state of the system, 
and the sum is over all possible states.

In our case, the system of particles obtained after the complete evaporation
of the black hole has twofold uncertainties - (i) how many particles have been emitted is not predetermined, and (ii) if it is given that the number of emitted particles is known, the particular microstate in which the black hole evaporated is not known. Both of these uncertainties
are there in the total entropy of the radiation system. This is illustrated below.

The probability that the black hole will emit $n$ particles before complete evaporation is given by Eqn.(\ref{q_n}). So, the first uncertainty is represented by a form of entropy denoted by $S(number)$ as
\begin{eqnarray}
S(number) &=& - \sum_{n=1}^{\infty} q_n \ln q_n \nonumber\\
&=& \sum_{n=1}^{\infty} \frac{\Omega_n}{\Omega_{total}} \ln 
\Omega_{total} - \sum_{n=1}^{\infty} \frac{\Omega_n}{\Omega_{total}} \ln 
\Omega_n
\nonumber \\
&=& \frac{\ln \Omega_{total}}{\Omega_{total}} \sum_{n=1}^{\infty} 
\Omega_n - \frac{1}{\Omega_{total}}\sum_{n=1}^{\infty}
\Omega_n \ln \Omega_n \nonumber \\
&=& \ln \Omega_{total} - \frac{\sum_{n=1}^{\infty}
\Omega_n \ln \Omega_n}{\sum_{n=1}^{\infty} \Omega_n}
\end{eqnarray}

Now, consider that the black hole is evaporated by emission of $n$ particles. The first particle is 
emitted with energy $E_1$, second particle with energy $E_2$, and the 
$n^{th}$ particle with energy $E_n$. The joint probability for this 
sequence of emission is
\begin{align}
P = &\Gamma(E_1;M) \times \Gamma(E_2;M-E_1) \times \ldots\nonumber\\
&\times \Gamma\left(E_n;M-\sum_{j=1}^{n-1}E_j\right) \nonumber \\
= &\exp\left(-4 \pi GM^2\right).
\end{align}
It is readily seen that this probability holds true for any value
of $n$, and for any partition of the total energy $M$ among the emitted particles.
That is to say, all $\Omega_{total}$ ways of black hole evaporation occur with this same probability. Since all $\Omega_{total}$ possibilities
occur with equal probability, it easily follows that
\begin{equation}
P=\frac{1}{\Omega_{total}}=e^{-4 \pi GM^2},
\end{equation}
reconfirming Eqn.(\ref{omegatotal}). Again, the probability of occurrence of one of the $\Omega_{total}$ possibilities of black hole evaporation can also be written as
\begin{align}
    P=&\text{Probability that the black hole has emitted}\nonumber\\
    &\text{$n$ particles} \times \text{Probability of occurrence of}\nonumber\\
    &\text{one among $\Omega_n$ microstates} \nonumber\\
    =& q_n \times P_{\alpha}(microstate|number),
\end{align}
where $P_{\alpha}(microstate|number), \alpha=1,2,\ldots,\Omega_n$ is the conditional probability of the occurrence of the $\alpha^{th}$ microstate among the $\Omega_n$ possible microstates provided that the black hole has been evaporated by the emission of $n$ particles
\footnote{An analogy may help. The probability of choosing a particular card, say the Queen of Hearts, from a deck of 52 cards is $\frac{1}{52}$. This probability can also be expressed as probability of choosing the set of Hearts($\frac{1}{4}$) $\times$ probability of choosing the Queen card provided that the set of Heart is chosen($\frac{1}{13}$).}. Therefore,
\begin{align}
    \frac{1}{\Omega_{total}} = \frac{\Omega_n}{\Omega_{total}}\times P_{\alpha}(microstate|number), \nonumber\\
    P_{\alpha}(microstate|number) = \frac{1}{\Omega_n}.
\end{align}

The second uncertainty about the specific microstate, given the knowledge of the number of particles emitted, is represented by another form of entropy denoted by $S(microstate|number)$. For the emission of $n$
particles, this entropy is expressed as
\begin{align}
    S_{n}(microstate&|number) \nonumber \\
    =-\sum_{\alpha=1}^{\Omega_n}
    &P_{\alpha}(microstate|number)\nonumber\\
    &\times\ln P_{\alpha}(microstate|number) \nonumber \\
=\sum_{\alpha=1}^{\Omega_n} &\frac{1}{\Omega_n} \ln\Omega_n \nonumber \\
=\ln\Omega_n&,
\end{align}

Finally, the total uncertainty about the particular way among the $\Omega_{total}$ possibilities in which the black hole
has evaporated encompasses both of the uncertainties that we discussed earlier. This total uncertainty is represented by the total entropy of the system as
\begin{eqnarray}
S_{rad} &=& -\sum_{k=1}^{\Omega_{total}} \frac{1}{\Omega_{total}} \ln 
\frac{1}{\Omega_{total}} \nonumber \\
&=& \ln \Omega_{total}=4\pi GM^2 \nonumber\\
&=& S_{BH}.
\end{eqnarray}

So,
\begin{align}
S(number)&= S_{BH}-\frac{\sum_{n=1}^{\infty} \Omega_n S_{n}(microstate|
number)}
{\sum_{n=1}^{\infty} \Omega_n}; \nonumber \\
S_{BH}&=S(number)+S_{avg}(microstate|number),\nonumber\\
&=S_{rad},
\end{align}
where $S_{avg}(microstate|number)$ is the conditional entropy $S(microstate|number)$ averaged over all possible numbers of emitted particles. So, we see that the total entropy of the system ($S_{rad}=S_{BH}$) contains two
parts:\\
(i) Entropy due to uncertainty in the number of particles emitted ($S(number)$),\\
(ii) Average entropy due to uncertainty in the microstates of a given number of emitted particles ($S_{avg}(microstate|number)$).

\section{\label{4} Summary and Conclusions}
We have revisited the process of non-thermal radiation from black holes.
The non-thermal correction to Hawking's original calculations of 
black hole radiation
comes from taking into account the backreaction of the emitted
particles on the black hole spacetime. Entropy is conserved during
the evaporation process of black holes due to non-thermal radiation,
irrespective of the number of particles emitted. In this work
we have tried to answer the following question: What is the probability that a
black hole of some given mass emits a certain number of particles before
being completely evaporated?
Conservation of entropy during the evaporation process plays a crucial
role in determining the probability.
We have found that a black hole of mass $M$ evaporates completely by
emitting $n$ particles in $\Omega_n=\frac{(4\pi GM^2)^{n-1}}{(n-1)!}$
different possible ways. These different possible ways correspond to different
possible partitionings of the Bekenstein-Hawking entropy of the initial
black hole among the emitted particles. If we let the number of emitted
particles be arbitrary, then there are in total $\Omega_{total}=e^{-4\pi GM^2}$
ways in which the black hole can evaporate completely. From these, we find
the probability of the emission of $n$ particles before complete black hole
evaporation to be $q_n=\frac{(4\pi GM^2)^{n-1}}{(n-1)!}e^{4\pi GM^2}$.

From the probability distribution obtained for different numbers of emitted
particles, we find that for black holes of mass $M$, there is a \textit{most probable number}
of emitted particles, $n_{max}=4\pi GM^2$. That is, the black hole is most
likely to evaporate completely by emitting $n_{max}$
particles. This implies that, for more massive black holes, larger
number of particles is expected to be emitted before their complete evaporation.
We have expressed this conclusion in the form of a displacement relation
between the most probable number of particles emitted and the temperature 
of the initial black hole, $n_{max}T_{BH}^2=\frac{1}{16\pi G}=\text{constant}$.
This displacement relation resembles Wien's displacement law for blackbody radiations.

Finally, we have examined the entropy of the system of radiated
particles obtained after complete evaporation of a black hole, from a different perspective.
As mentioned earlier, this entropy matches the entropy of the initial black hole.
When the entropy is interpreted as a measure of uncertainty
(or, equivalently hidden information), then we see that the total entropy
of the system contains two parts. One part contains the information about
the number of the particles in the system (the number of particles emitted
before the complete evaporation of the black hole). The other part contains
the information about the particular way in which the black hole has emitted
the given number of particles. This interpretation may give a new meaning
to the black hole entropy.
\begin{acknowledgements}
P.G. would like to thanks Department of Science $\&$ Technology, Govt. of India for financial support and Abhishek Atreya, Somshubhro Bandyopadhyay, and Koushik Ray for helpful discussions.
\end{acknowledgements}
\bibliography{ref}
\appendix
\section{\label{A} Surface area of the microstate hyperplane \& Area of
a unit cell in the microstate hyperplane}

\noindent
{Surface area of the microstate hyperplane:}
In this section we will calculate the surface area of an ($n-1$)
dimensional hyperplane given by the equation
\begin{equation}
\label{hyp}
\sum_{i=1}^{n} x_i = C,
\end{equation}
in the positive sector of the coordinates ($x_i\geq 0, \forall
i=1,2,..,n$).

\paragraph*{}
But first lets calculate the volume of the region bounded by this plane
and the planes $x_i=0,\forall i=1,2,..,n$. The n dimensional volume is
given by,
\begin{eqnarray}
\mathcal{V}_{n} &=& \int_{x_1=0}^{C}\int_{x_2=0}^{C-x_1}...
\int_{x_n=0}^{C-\sum_{i=1}^{n-1}x_i} \prod_{i=1}^{n}dx_i \nonumber \\
&=& \int_{x_1=0}^{C}... \int_{x_{n-1}=0}^{C-\sum_{i=1}^{n-2}x_i}
(C-\sum_{i=1}^{n-1}x_i) \prod_{i=1}^{n-1}dx_i \nonumber 
\end{eqnarray}
Let,
\begin{equation}
C-\sum_{i=1}^{n-2}x_i = \alpha.
\end{equation}
Therefore,
\begin{eqnarray}
\mathcal{V}_{n}&=&\int_{x_1=0}^{C}... \int_{x_{n-1}=0}^{\alpha}
(\alpha - x_{n-1}) \prod_{i=1}^{n-1}dx_i \nonumber \\
&=& \frac{1}{2} \int_{x_1=0}^{C}...
\int_{x_{n-2}=0}^{C-\sum_{i=1}^{n-3}x_i}
(C-\sum_{i=1}^{n-2}x_i)^2 \prod_{i=1}^{n-2}dx_i \nonumber
\end{eqnarray}
Again let,
\begin{equation}
C-\sum_{i=1}^{n-3}x_i = \beta.
\end{equation}
Therefore,
\begin{eqnarray}
\mathcal{V}_{n}&=&\int_{x_1=0}^{C}... \int_{x_{n-2}=0}^{\beta}
 (\beta - x_{n-2})^2 \prod_{i=1}^{n-2}dx_i \nonumber \\
&=& \frac{1}{2\times 3} \int_{x_1=0}^{C}...
\int_{x_{n-3}=0}^{C-\sum_{i=1}^{n-4}x_i}
(C-\sum_{i=1}^{n-3}x_i)^3 \prod_{i=1}^{n-3}dx_i \nonumber
\end{eqnarray}
Let,
\begin{equation}
C-\sum_{i=1}^{n-4}x_i = \delta.
\end{equation}
Therefore,
\begin{eqnarray}
\mathcal{V}_{n}&=&\int_{x_1=0}^{C}... \int_{x_{n-3}=0}^{\delta}
(\delta - x_{n-3})^3 \prod_{i=1}^{n-3}dx_i \nonumber \\
&=& \frac{1}{4 \times 3!} \int_{x_1=0}^{C}...
\int_{x_{n-4}=0}^{C-\sum_{i=1}^{n-5}x_i}
(C-\sum_{i=1}^{n-4}x_i)^4 \prod_{i=1}^{n-4}dx_i \nonumber
\end{eqnarray}
Proceeding in similar way, we get
\begin{eqnarray}
\mathcal{V}_{n}&=&\frac{1}{(n-2)!}\int_{x_1=0}^{C}
\int_{x_2=0}^{C-x_1}(C-x_1-x_2)^{n-2} dx_2 dx_1 \nonumber \\
&=&\frac{1}{(n-1)!}\int_{x_1=0}^{C}(C-x_1)^{n-1}dx_1 \nonumber \\
&=& \frac{1}{n!} C^n
\end{eqnarray}
Now, we turn back to calculate the surface area of the hyperplane
represented by the eqn. \ref{hyp}. In general, the surface area of any
($n-1$) dimensional surface embedded in $n$ dimensions, represented by
eqn.
\begin{equation}
x_n=\phi (x_1,x_2,...,x_{n-1})
\end{equation}
can be written as,
\begin{equation}
\mathcal{S}_{n-1} = \int_\mathcal{D} \sqrt{1+\sum_{i=1}^{n-1}
(\frac{\partial \phi}{\partial x_i})^2} \prod_{i=1}^{n-1} dx_i ,
\end{equation}
where, $\mathcal{D}$ is the projection of the hyperplane on the space
spanned by ($x_1,x_2,...,x_{n-1}$).

\paragraph*{}
For our case, the equation of the hyperplane can be written as
\begin{equation}
x_n=\phi=C-\sum_{i=1}^{n-1} x_i.
\end{equation}
So,
\begin{equation}
\frac{\partial \phi}{\partial x_i} = -1, \forall i=1,2,...,n-1.\nonumber
\end{equation}
Therefore,
\begin{equation}
\label{area}
\mathcal{S}_{n-1} = \int_\mathcal{D} \sqrt{n} \prod_{i=1}^{n-1} dx_i 
\end{equation}
Here, the projected region $\mathcal{D}$ on the space spanned by
($x_1,x_2,...,x_{n-1}$) is bounded by ($n-2$) dimensional surfaces
represented by eqns. $x_i =0, \forall i=1,2,..,n-1$ (since we are only
concerned about the positive sector of the space) and
$C-\sum_{i=1}^{n-1} x_i=0$. So the term $\int_\mathcal{D}
\prod_{i=1}^{n-1} dx_i$ is nothing but the volume of the ($n-1$)
dimensional space bounded by the said ($n-2$) dimensional surfaces. Now,
the surface area can be written as,
\begin{eqnarray}
\mathcal{S}_{n-1} &=& \sqrt{n} \int_{x_1=0}^{C}
\int_{x_2=0}^{C-x_1}...\int_{x_{n-1}=0}^{C-\sum_{i=1}^{n-2} x_i}
\prod_{i=1}^{n-1} dx_i \nonumber \\
&=& \sqrt{n} \mathcal{V}_{n-1} \nonumber \\
&=& \sqrt{n} \frac{C^{n-1}}{(n-1)!}.
\end{eqnarray}
This gives the surface area of the hyperplane. The area of the
microstate plane, whose equation is given by $\sum_{i=1}^{n} S_i =
S_{BH} $ can be calculated using this formula in a straight forward way.

\noindent
{Area of the unit cell:}
The area of a unit cell in the microstate hyperplane can be calculated
using eqn. \ref{area}, where the region $\mathcal{D}$ is the projection
of the unit cell on the ($n-1$) dimensional space spanned by
($S_1,S_2,..,S_{n-1}$). The projected region
$\mathcal{D}$ forms a ($n-1$) dimensional unit cube, then the area of
the unit cell in the microstate hyperplane is given by
\begin{equation}
\mathcal{A}_{n-1} = \sqrt{n}.
\end{equation}
The unit cell in the microstate hyperplane and the projected region
$\mathcal{D}$ are shown in fig. \ref{projection} for the case when the
black hole is evaporated by 3 particles.

\section{\label{B} Value of the most probable number of emitted
particles ($n_{max}$) from the distribution fuction}

In this section we will evaluate the most probable number of emitted
particles ($n_{max}$) analytically from the probabaility distribution
function. We have the probability distribution function as
\begin{equation}
q_n = e^{-S_{BH}}\frac{S_{BH}^{n-1}}{(n-1)!}.
\end{equation}
For the most probable number of emitted particles,
\begin{equation}
q_{n_{max}}=\text{maximum}. \nonumber
\end{equation}
Now, $q_n$ is a function of discrete variable. To obtain the maxima for
this function, the following 2 conditions has to be satisfied
simultaneously.
\begin{eqnarray}
q_{(n_{max}+1)}-q_{(n_{max})} \leq 0, \nonumber \\
q_{(n_{max})}-q_{(n_{max}-1)} \geq 0.
\end{eqnarray} 
Therefore,
\begin{eqnarray}
e^{-S_{BH}}\left[\frac{S_{BH}^{(n_{max})}}{(n_{max})!}-
\frac{S_{BH}^{(n_{max}-1)}}{(n_{max}-1)!}\right] &\leq & 0 \nonumber \\
\frac{S_{BH}^{(n_{max})}}{(n_{max})!}-
\frac{n_{max}S_{BH}^{(n_{max}-1)}}{(n_{max})!} &\leq & 0 \nonumber \\
S_{BH} \leq n_{max}
\end{eqnarray}
and,
\begin{eqnarray}
e^{-S_{BH}}\left[\frac{S_{BH}^{(n_{max}-1)}}{(n_{max}-1)!}-
\frac{S_{BH}^{(n_{max}-2)}}{(n_{max}-2)!}\right] &\geq & 0 \nonumber \\
\frac{S_{BH}^{(n_{max}-1)}}{(n_{max}-1)!}-
\frac{(n_{max}-1)S_{BH}^{(n_{max}-2)}}{(n_{max}-1)!}
&\geq & 0 \nonumber \\
S_{BH}+1 \geq n_{max}.
\end{eqnarray}
So for the given probability distribution, $n_{max}$ has the integer
value in the range
\begin{equation}
S_{BH} \leq n_{max} \leq S_{BH}+1.
\end{equation}
\end{document}